\documentclass[pop,amsmath,amssymb,twocolumn,superscriptaddress,floatfix]{revtex4}

\usepackage{color}

\usepackage{amsmath}
\usepackage{amssymb}
\usepackage{amsbsy}
\usepackage{amsfonts}
\usepackage{amstext}   
\usepackage{array}    
\usepackage{graphicx}
\usepackage{xspace}

\newcommand{\gmax}{\ensuremath{g_\textrm{max}}\xspace}
\newcommand{\smax}{\ensuremath{S_\textrm{max}}\xspace}
\newcommand{\rh}{\ensuremath{r_{1/2}}\xspace}
\newcommand{\geff}{\ensuremath{\Gamma^\textrm{eff}}\xspace}
\renewcommand{\vec}{\mathbf}
\renewcommand{\vec}[1]{\boldsymbol{#1}}

\begin{document}

 \author{T. Ott}
 \affiliation{%
     Christian-Albrechts-Universit\"at zu Kiel, Institut f\"ur Theoretische Physik und Astrophysik, Leibnizstra\ss{}e 15, 24098 Kiel, Germany
 }%

 \author{M. Bonitz}%
 \affiliation{%
     Christian-Albrechts-Universit\"at zu Kiel, Institut f\"ur Theoretische Physik und Astrophysik, Leibnizstra\ss{}e 15, 24098 Kiel, Germany
 }%

 \author{L. Stanton}%
 \affiliation{%
Lawrence Livermore National Laboratory, Livermore, CA 94550, USA}%

 \author{M. S. Murillo}%
 \affiliation{%
Los Alamos National Laboratory, Los Alamos, NM 87545, USA }%

\title{{Coupling Strength in Coulomb and Yukawa One-Component Plasmas}}

\date{\today}

\begin{abstract}

In a non-ideal classical Coulomb one-component plasma (OCP) all thermodynamic properties are known to depend only on a single parameter -- the coupling parameter $\Gamma$. In contrast, if the pair interaction is screened by background charges (Yukawa OCP) the thermodynamic state depends, in addition, on the range of the interaction via the screening parameter $\kappa$. How to determine in this case an effective coupling parameter has been a matter of intensive debate.
Here we propose a consistent approach for defining and measuring the coupling strength in Coulomb and Yukawa OCPs based on a fundamental structural quantity, the radial pair distribution function (RPDF). The RPDF is often accessible in experiments by direct observation or indirectly through the static structure factor. Alternatively, it is directly computed in theoretical models or simulations. Our approach is based on the observation that the build-up of correlation from a weakly coupled system proceeds in two steps: First, a monotonically increasing volume around each particle becomes devoid of other particles (correlation hole), and second (upon further increase of the coupling), a shell structure emerges around each particle giving rise to growing peaks of the RPDF. Using molecular dynamics simulation, we present a systematic study for the dependence of these features of the RPDF on $\Gamma$ and $\kappa$ and derive a simple expression for the effective coupling parameter.
\end{abstract}

 \maketitle

\section{Introduction}
Strongly coupled or strongly correlated systems are abundant in many fields of physics. In plasma physics, they have come into the focus in recent times due 
to the increasing availability of experimental realizations. In these experiments, the mutual repulsion of the like-charged particles is comparable to
or even greater than their thermal kinetic agitation. This leads to the emergence of complex phenomena such as shear waves~\cite{Liu2010a,Ott2013}, solidification~\cite{Hartmann2010}, 
cooperative behavior~\cite{Coueedel2014} 
and anomalous transport~\cite{Ott2009b,Feng2010}. Because of these rich physics, there is strong aspiration to reach ever stronger degrees of correlation in the experiments.

However, despite the central role of the coupling strength, it is often difficult to assess from experimental data as it requires detailed knowledge about the system state. 
In addition, the role of Debye-screened interaction is often neglected when statements about the coupling strength are made. This leads to difficulties in comparing 
the degree of correlation across experiments which include dusty plasmas~\cite{Bonitz2010a}, ultracold neutral plasmas~\cite{Killian2007}, ions in traps~\cite{Dantan2010}, and warm dense matter setups. 

The situation is even more complex in a system containing multiple species such as two-component plasmas. Here, in principle, one has to distinguish the coupling strength of the two components as well as the inter-species coupling. In high density plasmas, such as warm dense matter, where the light component (i.e., the electrons) may be quantum degenerate and weakly coupled, the heavy component may be classical and strongly coupled. In addition, partial ionization maybe relevant making the analysis of structural and thermodynamic properties challenging. 

In this work, we give a structrual definition of the coupling strengths in nonideal plasmas, focusing exclusively on one-component plasmas. 
There, the coupling strength is typically given in terms of a coupling parameter $g$ of the form 
\begin{align}
 g &= \frac{\langle E_\textrm{NN}\rangle}{k_BT}, \label{eq:coupling_g}
\end{align}
where $E_\textrm{NN}$ is a measure of the typical nearest-neighbor interaction. Strong coupling is associated with $g>1$. For Coulomb systems with the interaction potential $V(r)=Q/r$, 
the coupling takes the form
\begin{align}
 \Gamma=\frac{Q^2/a}{k_BT}, \label{eq:coulomb_g}
\end{align}
where $Q$ is the particle charge and $a=\left[3/\left (4 n \pi \right)  \right]^{1/3}$ is the Wigner-Seitz radius, a measure of the nearest-neighbor distance. 
Evaluation of $\Gamma$ for a given experimental situation thus requires the measuring of $T$ and $Q$ separately (alongside the density $n$). 
The importance of the coupling parameter for a model one-component Coulomb plasma (OCP) arise from the observation that its mean energy and all thermodynamic quantities do not depend on density and temperature separately but only via $\Gamma$.

In Yukawa systems, the interaction potential takes the form $V(r)=Q/r \times \exp(-\kappa r/a)$, where $\kappa$ is related to the inverse of the screening length $\lambda$ by $\kappa=a/\lambda$. In a classical plasma $\lambda$ is given by the Debye length whereas, in a strongly degenerate quantum plasma, it is given by the Thomas-Fermi screening length. In general, $\lambda$ is defined by the static long-wavelength limit of the longitudinal polarization function, see, e.g.,  Ref.~\cite{Bonitz1998}. 
It is customary to give the coupling strength of Yukawa systems in terms of $\Gamma$ as per Eq.~\eqref{eq:coulomb_g} together with the inverse screening length $\kappa$. 
However, since $Q^2/a$ is not the true measure of the potential energy of a Yukawa system (due to the missing screening~\cite{Lyon2013}), $\Gamma$ carries no immediate physical significance for 
these systems. This raises the question of how to compare Yukawa and Coulomb systems on the one hand and Yukawa systems of different screening on the other. 

We thus face two intricacies when using $\Gamma$ as a measure of the coupling strength: 1) It requires the measurement of $Q$ and $T$ separately, and 2) it is only 
physically meaningful for Coulomb systems. The goal of this work is to alleviate these problems by finding a one-to-one mapping between the structure of Coulomb and Yukawa systems 
to~$\Gamma$ [Eq.~\eqref{eq:coulomb_g}]. This allows one to infer $\Gamma$ from the structure alone, i.e., without knowledge of $Q$ or $T$. Furthermore, by using 
structural features to define the coupling strength, it is possible to give an effective coupling parameter \geff for Yukawa systems~\cite{Murillo2008, Ott2011b, Hartmann2005}. For a Yukawa system 
with a given screening length, this parameters equals the value of $\Gamma$ of the corresponding Coulomb system with the most similar structure, i.e., the most meaningful comparison system. A similar 
approach as been used in Ref.~\cite{Clerouin2013}. 

The path taken here towards this goal is the following: Using Langevin Dynamics simulation, we obtain reference data for the structure of Coulomb and Yukawa systems in the form of the radial pair distribution function (RPDF) $g(r)$. To uniquely relate the shape of this function to the physical degree of non-ideality we use two of its properties: the width of the correlation hole and the height of the first peak of $g(r)$. 

Comparing these experimentally accessible quantities to the reference data allows one to infer $\Gamma$. Carefully optimized fit formulas are derived 
which connect the structure to the coupling strength and allow for an interpolation to values not covered in the reference data. 

\begin{figure}
   \begin{center}\includegraphics[scale=1]{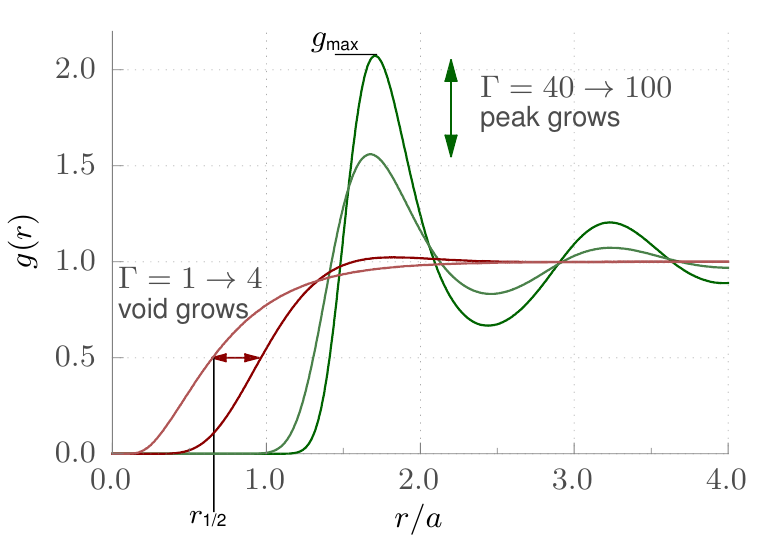}\end{center}
   \caption{Radial pair distribution function of a Coulomb system at $\Gamma=1;4;40;100$. \label{fig:gofr_visual}}
\end{figure}

\section{Methods and Simulation}
We use Langevin dynamics simulation for $N=8192$ particles to obtain the equilibrium properties of the Coulomb and Yukawa One-Component plasma. The Langevin equation reads
\begin{align}
m \ddot{{\vec r}}_i =\vec F_i - m \bar \nu \vec v_i + \vec y_i \, \qquad i=1\dots N\, \label{eq:langevin} 
\end{align}
where $\bar \nu$ is the friction coefficient, $\vec F_i$ is the repulsive interaction force between the particles and
$\vec y_i(t)$ is a Gaussian white noise with zero mean and the standard deviation
$
   \langle y_{\alpha,i}(t_0)y_{\beta,j}(t_0+t)\rangle=2k_BTm\bar \nu\delta_{ij}\delta_{\alpha\beta}\,\delta(t)
$ [$\alpha,\beta\in \{x,y,z\}$]. We use a constant friction coefficient of $\bar \nu = 0.1\omega_p$, where $\omega_p= [ 4\pi Q^2 n/m]^{1/2}$ is the nominal (Coulomb) plasma frequency. 
We vary $\kappa$ between $0$ and $2$ in steps of $0.2$ and vary $\Gamma$ to cover the entire liquid phase (i.e., from $\Gamma=1$ up to close to the phase transition temperature). 

To assess the structural state of the system, we use the radial pair distribution defined as 
\begin{align}
g(\vec r) = \frac{1}{Nn} \left\langle \sum_{i,j=1 \atop i \neq j}^N \delta(\vec r-  \vec r_{ij}) \right\rangle, \label{eq:pdf}
\end{align}
where $r_{ij}=\vert \vec r_i - \vec r_j\vert$ and the averaging is over time.  The RPDF is the most simple structural 
quantity of a many-particle system and describes the relative occurrence frequency of a particular pair distance $r$ in the system. In many setups, it is experimentally accessible through direct optical monitoring (e.g., in dusty plasmas) or indirectly through the measurement of the static structure factor (e.g., through scattering measurements). There are also various theoretical approaches to calculate the RPDF including 
the Hypernetted Chain Approach~\cite{Ng1974,Bruhn2011} and simulations~\cite{Brush1966}. 

For a system of non-interacting particles, $g(r)\equiv 1$ and any correlation effects will manifest themselves in deviation of $g(r)$ from unity. 
One of the two main RPDF characteristics of strongly coupled  
systems is the correlation void at small values of $r$ which reflects the mutual repulsion of particles at small distances. The second feature is the emergence of a series of peaks in $g(r)$ related 
to the formation of shell-like structures of first, second, etc. neighbors around any particular particle.

\begin{figure}
   \begin{center}\includegraphics[scale=1]{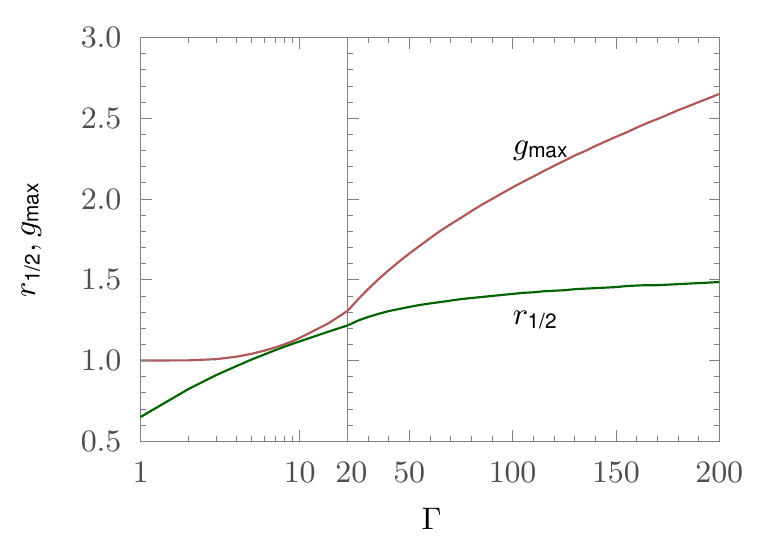}\end{center}
   \caption{\gmax and \rh as a function of $\Gamma$ for a Coulomb system. Note that in the left part of the figure, the $\Gamma$-axis is scaled logarithmically.  \label{fig:gmaxr05}}
\end{figure}
 
The build-up of correlation from an uncorrelated system manifests itself in two subsequent steps (Fig.~\ref{fig:gofr_visual}): First, the correlation void grows rapidly as the 
correlation increases. After this process, upon further increase of the coupling strength, the shell structure emerges and becomes gradually more pronounced. Notably, 
the size increase of the correlation void with the coupling strength is rapid only at small values of the coupling, while the growth of the peak structure is most prominent at larger coupling values. 
This complementary development leads us to use both features of the RPDF in our subsequent analysis (see Fig.~\ref{fig:gofr_visual}). Specifically, we assess the size of the correlation void as the value of $\rh$ defined 
by
\begin{align}
g(a\cdot \rh) = 0.5, 
\end{align}
i.e., the dimensionless distance at which $g(r)$ has risen to half its asymptotic limit. For the peak structure, we use the height of the first peak in the RPDF, i.e., its 
global maximum \gmax~\footnote{Additional properties of the RPDF will be the subject of a forthcoming paper.}. 

As an example, we show in Fig.~\ref{fig:gmaxr05} both \rh and \gmax as a function of $\Gamma$ for a Coulomb system (note that in the left 
part of Fig.~\ref{fig:gmaxr05}, the $\Gamma$-axis is scaled logarithmically). At small~$\Gamma$, \rh varies rapidly and \gmax varies only slowly while for larger $\Gamma$, the inverse is true. 
However, since both \rh and \gmax are strictly monotonic functions of $\Gamma$ at given $\kappa$, they both uniquely define the whole structural composition of the system and thus the complete thermodynamic equilibrium properties of the plasma. The switching between \rh and \gmax to define the coupling strength is thus only a matter of practicality. 

\section{Structure and Coupling Strength}
We now establish a one-to-one mapping between the structure of a system and its physical coupling strength. In doing so, we assume that the screening length $\kappa$ 
is known in a given experimental setup. This is crucial, because in this first step, we express the coupling of a Yukawa system in the customary way through the nominal Coulomb coupling parameter $\Gamma$. Since this is not the actual physical coupling strength of a Yukawa system, systems with the same value of $\Gamma$ but different $\kappa$ exhibit different degrees of structural correlation. Conversely, a given structure can correspond to a multitude of $\{\Gamma, \kappa\}$ pairs, so that without knowledge of $\kappa$, these systems cannot be distinguished. If, however, $\kappa$ is known, then the structure [i.e. the shape of $g(r)$] uniquely defines the coupling value $\Gamma$. 

\begin{figure*}[p]
   \begin{center}\includegraphics[width=12cm]{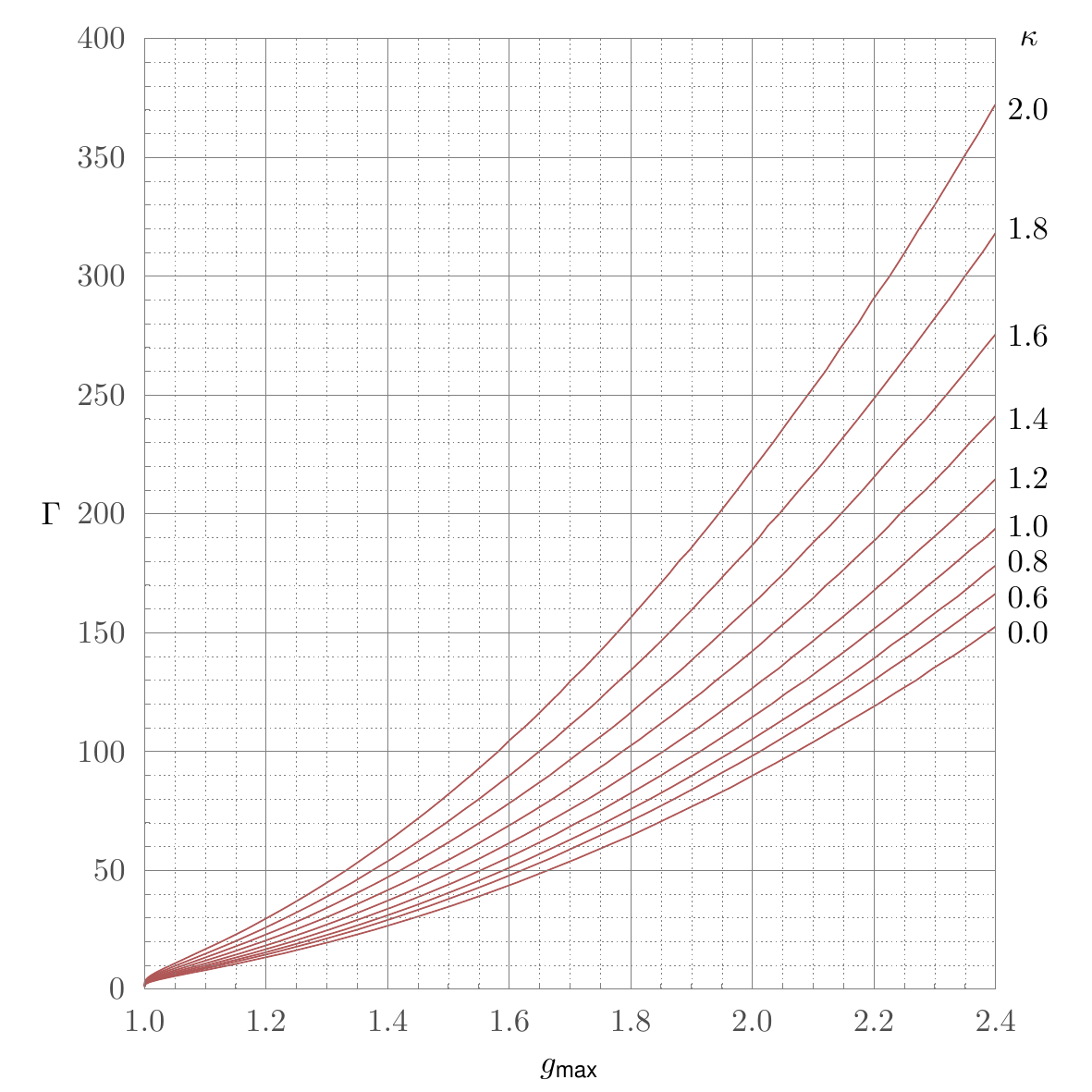}\end{center}
   \begin{center}\includegraphics[width=12cm]{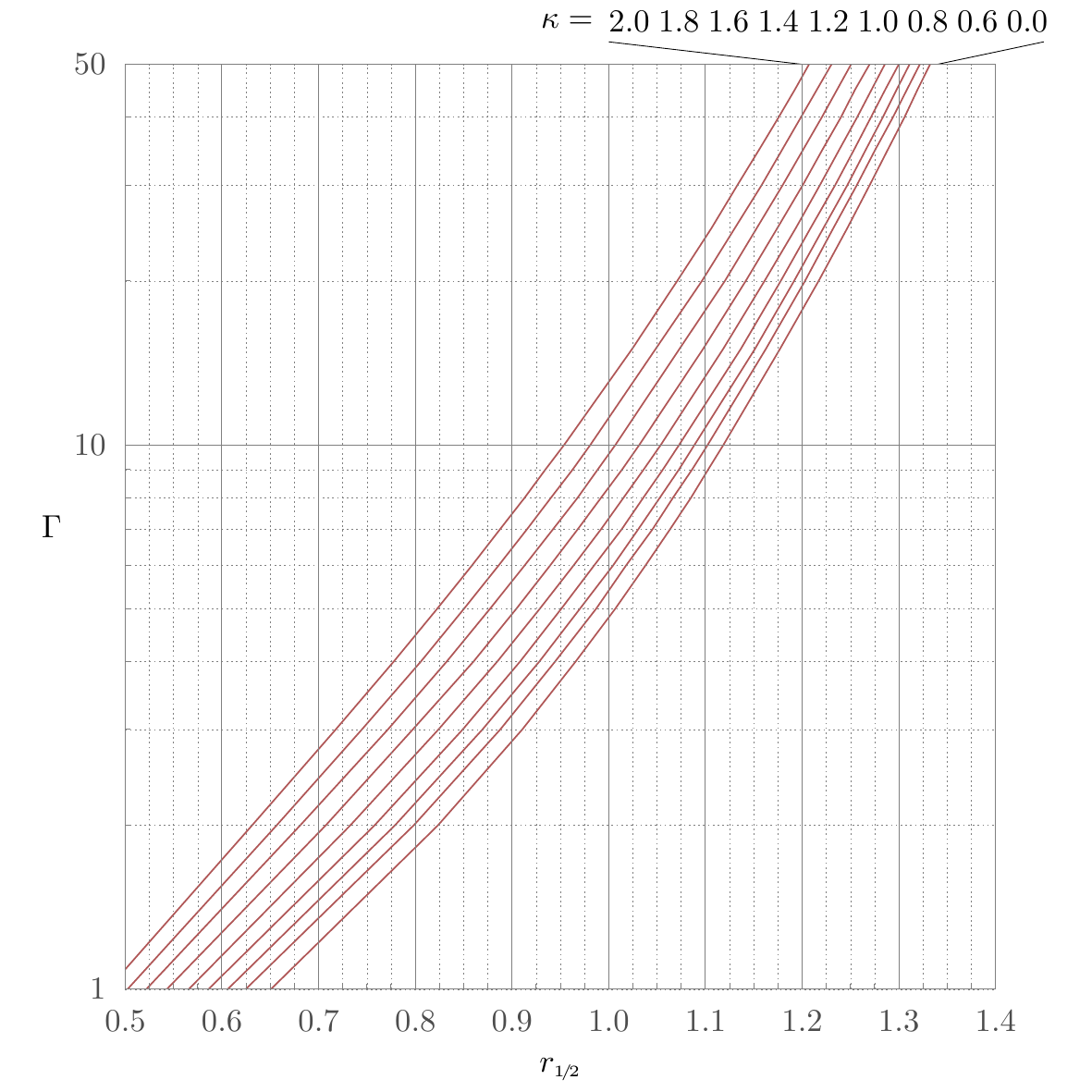}\end{center}
   \caption{\emph{Top:} Coupling parameter $\Gamma$ as a function of the peak height \gmax of the RPDF for $\kappa$ as indicated in the figure. 
            \emph{Bottom:} Coupling parameter $\Gamma$ as a function of $\rh$ (see text for definition) for $\kappa$ as indicated in the figure.  \label{fig:graphrel}}
\end{figure*}

First, we consider systems whose correlation is large enough that a peak structure has formed and the first peak in the RPDF is clearly visible (trivially, \gmax tends to unity when the coupling strength is lowered). The relation between \gmax and $\Gamma$ at a given $\kappa$ is shown in the top graph of Fig.~\ref{fig:graphrel}. 
The unique correspondence between $\Gamma$ and \gmax (at given $\kappa$) is clear from these data. One also sees that a given peak height can correspond to several 
values of $\Gamma$, depending on $\kappa$, or, conversely, that for a given $\Gamma$, systems with higher $\kappa$ have lower peak heights and are thus less strongly coupled. This is a 
reflection of the fact that in systems with large values of $\kappa$, the increased screening reduces the interaction between the particles. 
To use \gmax as a reliable indicator of $\Gamma$, the dependence $\Gamma(\gmax)$ must be sensitive to a variation of \gmax. This corresponds to those parts of the curves in Fig.~\ref{fig:graphrel}
which have a sufficiently high slope, i.e., $\gmax \gtrsim 1.4$. 

For less strongly coupled systems, $\Gamma$ is only a weak function of \gmax, since the peak develops slowly as the coupling is increased. Instead, the build-up 
of correlation at these small coupling values is indicated by the growth of the correlation void \rh. The dependence of $\Gamma$ on \rh is shown in the bottom graph of Fig.~\ref{fig:graphrel}. 
It is clear that \rh is a sensitive indicator of $\Gamma$ in the range $\rh\lesssim 1.3$ and can be used to deduce $\Gamma$ at a given $\kappa$. The complementary nature 
of \gmax and \rh is readily observable from the two graphs of Fig.~\ref{fig:graphrel}: While \gmax is sensitive at higher values of $\Gamma$, \rh is sensitive at small values of $\Gamma$. 
There is an overlap around $\Gamma\approx 30\ldots 50$ in which both methods are usable and give identical results. We stress that the use of both \gmax and \rh is made to maximize the 
sensitivity of the measurement during both stages of the correlation build-up. 

We now develop approximate fit formulas to the data of Fig.~\ref{fig:graphrel} in which the \emph{relative} mean-squared deviation is minimized~\cite{Schmidt2009}. 
For $\Gamma(\gmax)$, the data is well described by the polynomial 
\begin{align}
 \Gamma(g_\textrm{max},\kappa) &= a_1(\kappa) + a_2(\kappa) g_\textrm{max} + a_3(\kappa) g_\textrm{max}^2, \nonumber\\
 & \hspace{3.5cm}1.4<\gmax<2.4\label{eq:gamma_gmax}, 
\end{align}
where $a_i(\kappa)$ is given by the values in Table~\ref{tab:gmax_fit}. From the errors given in Table~\ref{tab:gmax_fit}, it is clear that Eq.~\eqref{eq:gamma_gmax} 
is an excellent fit to the data. To simplify the usage of Eq.~\eqref{eq:gamma_gmax} and interpolate to intermediate values of $\kappa$, the functional form of 
$a_i(\kappa)$ can be further approximated by
 \begin{align}
  a_1(\kappa) &= 22.40 - 7.88 \kappa + 9.68 \kappa^2 \nonumber\\
  a_2(\kappa) &= -70.09 + 20.28 \kappa - 32.48 \kappa^2 \label{eq:gamma_gmax_fit}\\
  a_3(\kappa) &= 52.60 - 12.71 \kappa + 23.73 \kappa^2 \nonumber.
 \end{align}
 
Use of Eqs.~\eqref{eq:gamma_gmax_fit} in \eqref{eq:gamma_gmax} yields an maximum error of $3.3\%$ and an average error of $1.45\%$ over all numerical data. 

\begin{table}
 \begin{tabular}{m{1cm}m{1cm}m{1.5cm}m{1cm}cc}
$\kappa$ & $a_1$     & $a_2$       & $a_3$      & $\Delta_\textrm{max}$ (\%) & $\Delta_\textrm{avg}$ (\%) \\ \hline
0.0   & 22.864 & -68.942  & 51.209  & 0.22         & 0.09         \\
0.2   & 18.640 & -64.525  & 50.289  & 0.20         & 0.06         \\
0.4   & 21.858 & -69.515  & 52.682  & 0.16         & 0.06         \\
0.6   & 22.704 & -72.805  & 55.285  & 0.22         & 0.07         \\
0.8   & 23.608 & -77.265  & 59.012  & 0.22         & 0.08         \\
1.0   & 23.765 & -82.112  & 63.732  & 0.24         & 0.07         \\
1.2   & 26.843 & -91.755  & 70.828  & 0.18         & 0.07         \\
1.4   & 28.297 & -101.537 & 79.289  & 0.29         & 0.08         \\
1.6   & 35.852 & -120.384 & 91.765  & 0.19         & 0.09         \\
1.8   & 37.093 & -135.352 & 105.181 & 0.27         & 0.07         \\
2.0   & 47.348 & -163.976 & 124.738 & 0.30         & 0.08         \\ \hline
\end{tabular}
\caption{Fit parameters of Eq.~\eqref{eq:gamma_gmax}. Also given are the maximum and average deviation from the numerical data. \label{tab:gmax_fit} }
\end{table}

The dependence $\Gamma(\rh)$ is approximated by the following relation, 
\begin{align}
  \Gamma(\rh, \kappa) &= b_1(\kappa) \exp\big({b_2 \rh^3}\big) + b_3(\kappa),\nonumber\\
  & \hspace{2.5cm}\Gamma\geq 1 \textrm{ ~and~ } \rh<1.3\label{eq:gamma_r05},
\end{align}
where $b_2=1.575$ is a constant and the other fit parameters are given in Table~\ref{tab:r05_fit} alongside the maximum and average deviation from the numerical data. 
For $b_i$, the following approximation can be given: 
\begin{align}
  b_1(\kappa) &= 1.238 - 0.280 \kappa + 0.644 \kappa^2 \nonumber\\
  b_2\phantom{(\kappa)} &= 1.575 \label{eq:gamma_r05_fit}\\
  b_3(\kappa) &= -0.931 + 0.422 \kappa - 0.696 \kappa^2 \nonumber,
\end{align}  
which, in combination with Eq.~\eqref{eq:gamma_r05}, give a maximum error of $5.59$\% and an average error of $1.68$\% over all numerical data. 
 
\begin{table}[t]
\begin{tabular}{@{}m{1cm}m{1cm}m{1cm}cc@{}}
$\kappa$ & $b_1$    & $b_3$     & $\Delta_\textrm{max}$ (\%) & $\Delta_\textrm{avg}$ (\%) \\ \hline
0.0   & 1.200 & -0.873 & 2.46         & 1.77         \\
0.2   & 1.211 & -0.882 & 2.80         & 1.65         \\
0.4   & 1.254 & -0.909 & 2.38         & 1.34        \\
0.6   & 1.336 & -0.978 & 2.01         & 1.06         \\
0.8   & 1.453 & -1.073 & 1.62         & 0.92         \\
1.0   & 1.608 & -1.210 & 1.15         & 0.51         \\
1.2   & 1.811 & -1.400 & 1.66         & 0.46         \\
1.4   & 2.071 & -1.651 & 1.63         & 0.54         \\
1.6   & 2.400 & -2.985 & 3.93         & 1.22         \\
1.8   & 2.800 & -2.405 & 3.61         & 1.03         \\
2.0   & 3.307 & -2.940 & 1.24         & 0.38         \\ \hline
\end{tabular}
\caption{Fit parameters for Eq.~\eqref{eq:gamma_r05}.  Also given are the maximum and average deviation from the numerical data. \label{tab:r05_fit}}
\end{table}

Thus, in conclusion, we have developed a measure of the coupling strength based solely on structural features of the system. From this follow two complementary 
methods to infer the value of $\Gamma$ from the structure of the system. Figure~\ref{fig:graphrel} shows the 
relationship between \gmax and $\Gamma$ and between \rh and $\Gamma$ and can be used directly to obtain $\Gamma$ from the RPDF at a given $\kappa$. Equations~\eqref{eq:gamma_gmax} 
and \eqref{eq:gamma_r05} provide a more convenient means and allow the interpolation to intermediate screening lengths while only incurring a small error. Together these methods 
provide a non-invasive measurement method for $\Gamma$ for both Coulomb and Yukawa systems. The only knowledge required is the density $n$ (to obtain $a$) and either the 
peak height \gmax or the correlation void size \rh, both of which are generally much easier to measure than the charge state $Q$ and the kinetic temperature $T$. 

\section{Effective coupling strength}
After having addressed the problem of the definition and measurement of the coupling parameter $\Gamma$ in the previous section, we now turn to the question of how a unified effective 
coupling parameter \geff can be defined which carries physical significance not only for Coulomb but also for Yukawa systems. This problem has been considered before, 
especially for two-dimensional Yukawa systems~\cite{Ott2011b,Ikezi1986,Totsuji2001,Vaulina2002a,Hartmann2005} but also for three-dimensional systems, based, e.g., on the 
packing fraction~\cite{Murillo2008}. 
\begin{figure}
\includegraphics[scale=1.0]{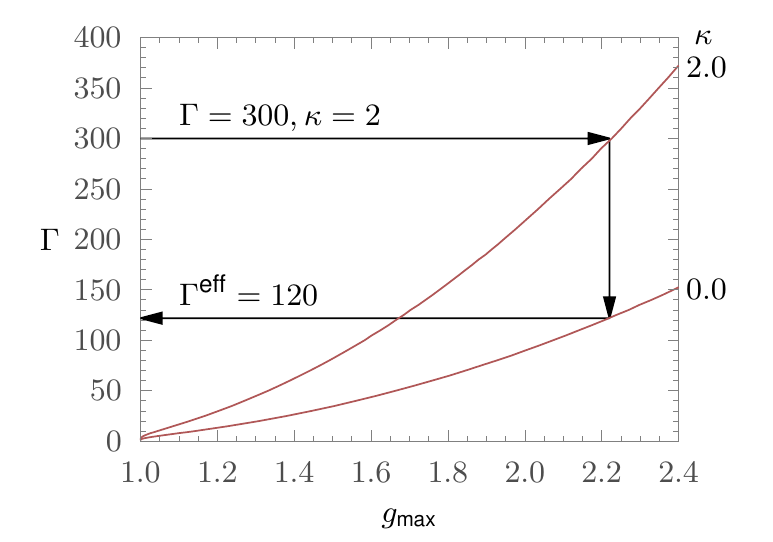}
\caption{Sketch of the graphical definition of \geff: A Yukawa system with $\Gamma=300$ and $\kappa=2$ has an 
effective structural coupling of $\geff=120$. \label{fig:gmax_sketch}}
\end{figure}

Here, our goal is to give a definition based on the structural information contained in \gmax and \rh, answering, in essence, the question ``Given a Yukawa system with a known 
nearest-neighbor correlation (i.e., a given \gmax or \rh), what is the structurally most similar Coulomb system?''

This question can be answered by a graphical solution based on Fig.~\ref{fig:graphrel} whose principle is sketched in Fig.~\ref{fig:gmax_sketch}: The 
value of \gmax or \rh for the known Yukawa system is projected down on the corresponding $\Gamma(\gmax)$ curve for $\kappa=0$ 
and the corresponding $\Gamma$, which is now equivalent to \geff, is read off. For situations in which both $\Gamma$ and $\kappa$ of a Yukawa 
system are known (e.g., in simulations), the corresponding \gmax or \rh can be obtained directly from Fig.~\ref{fig:graphrel} as well. 

For a formulaic solution to the question posed above, one needs to invert the relation $\Gamma(\gmax,\kappa)$ [Eq.~\eqref{eq:gamma_gmax}] to yield $\gmax(\Gamma,\kappa)$ and 
obtain \geff as 
\begin{align}
 \geff(\Gamma,\kappa) & \stackrel{.}{=} \Gamma(\gmax(\Gamma,\kappa),\kappa=0). \label{eq:geffgmax}
\end{align}

The same procedure yields, \emph{mutatis mutandis}, the complementary definition
\begin{align}
 \geff(\Gamma,\kappa) & \stackrel{.}{=} \Gamma(\rh(\Gamma,\kappa),\kappa=0). \label{eq:geffr05}
\end{align}

\begin{figure}
\includegraphics[scale=1.0]{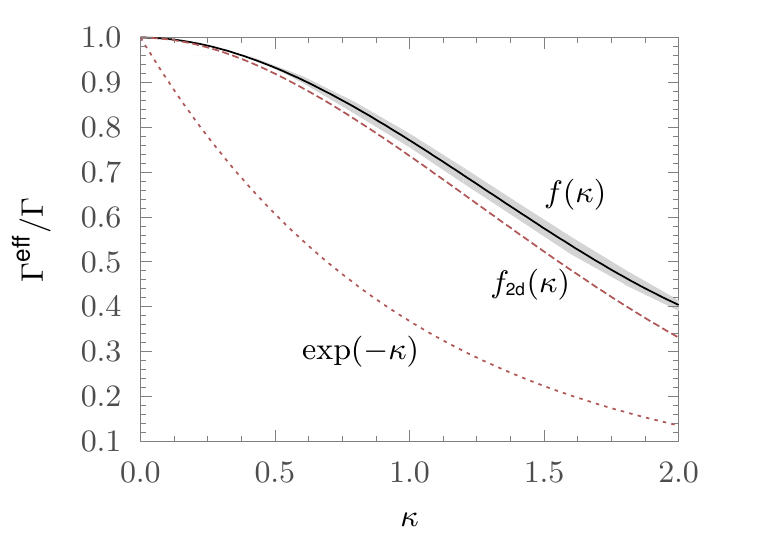}
\caption{The scaling function $f(\kappa)$~\eqref{eq:geff_scaling} as a function of $\kappa$. The bounds of the gray shaded area show the 
linearization of Eq.~\eqref{eq:geffgmax} (upper bound) and Eq.~\eqref{eq:geffr05} (lower bound). 
Also shown are the intuitive scaling function $\exp(-\kappa)$ 
and the scaling function for two-dimensional Yukawa systems~\cite{Hartmann2005}. \label{fig:geff_scaling}}
\end{figure}

Note that both Eq.~\eqref{eq:geffgmax} and \eqref{eq:geffr05} carry the same limitations for their application as Eqs.~\eqref{eq:gamma_gmax} [$1.4<\gmax<2.4$] 
and \eqref{eq:gamma_r05} [$\Gamma\geq 1 \textrm{ ~and~ } \rh<1.3$], respectively. Taken together, they provide a definition of \geff over a range of coupling strengths 
equivalent to a Coulomb system with $\Gamma=1\ldots 150$, i.e., over practically the whole strongly coupled liquid regime (crystallization of a Coulomb system 
occurs at $\Gamma=172$). 

A further simplification can be introduced by noticing that Eqs.~\eqref{eq:geffgmax} and \eqref{eq:geffr05} are very well approximated by their respective linearizations. 
In addition, for a given $\kappa$, the two linearizations of these equations coincide within 3\% with their joint average, which leads us to the following simple definition of \geff:
\begin{align}
 \geff(\Gamma,\kappa) &= f(\kappa) \cdot \Gamma, \hspace{1cm} 0\leq\kappa\leq 2,\nonumber\\ 
 & \hspace{2.8cm}1\leq\geff\leq 150 \label{eq:geff}
\end{align}
where the scaling function is given by
\begin{align}
 f(\kappa) &= 1 - 0.309 \kappa^2 + 0.0800 \kappa^3.\label{eq:geff_scaling}
\end{align}
The scaling function has been found as a least-square fit to the average of the linearizations. In this way, the definition \eqref{eq:geff} is the most accurate 
representation of $\geff(\Gamma,\kappa)$ valid for the whole liquid range in which $1<\geff<150$ since it incorporates both linearizations of~\eqref{eq:geffgmax} and 
\eqref{eq:geffr05}~\footnote{This is in contrast to the definitions of \geff for two-dimensional given in Refs.~\cite{Ott2011b, Hartmann2005} which only consider \gmax and are
thus only valid for more strongly coupled systems. }.

\begin{figure}[t]
\includegraphics[scale=1.0]{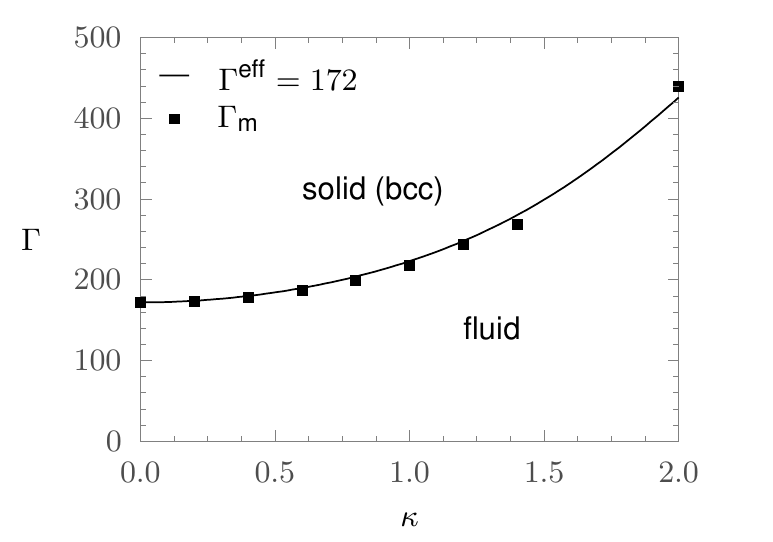}
\caption{$\kappa-\Gamma$ phase diagram for Yukawa systems. The symbols indicate the melting transition~\cite{Hamaguchi1997}. The solid line marks a constant effective coupling parameter $\geff=172$. Note that, for larger $\kappa$ the phase diagram is more complex due to the existence of an additional fcc-lattice phase (not shown)~\cite{Hamaguchi1997}.\label{fig:geff_melting}}
\end{figure}

The dependence of $f(\kappa)$ on $\kappa$ is shown in Fig.~\ref{fig:geff_scaling} together with the intuitive scaling function $\exp(-\kappa)$ which follows from 
straightforward application of Eq.~\eqref{eq:coupling_g} for a Yukawa system. Clearly, such a simple approach fails to capture the true structural coupling described 
by \geff. Figure~\ref{fig:geff_scaling} also shows the corresponding scaling function for a two-dimensional Yukawa system as derived by Hartmann~\emph{et al.}~\cite{Hartmann2005} which is valid 
for $\geff\gtrsim 40$ and where $\kappa=(\lambda\sqrt{\pi n})^{-1}$ is the two-dimensional definition of the screening strength. This comparison shows that the nominal screening of a Yukawa system 
has a stronger effect on the effective coupling in two dimensions than it does in three dimensions. 

Finally, we consider the liquid-solid phase transition of Coulomb and Yukawa fluids. Since this transition occurs when the the ratio of potential and kinetic energy 
exceeds a threshold value, one expects an effective structural coupling to be an indicator of the phase change. Figure~\ref{fig:geff_melting} shows the value $\Gamma_\textrm{m}$ 
at which the phase transition occurs~\cite{Hamaguchi1997} together with a constant effective coupling parameter of $\geff=172$ according to our definition~\eqref{eq:geff}. 
Evidently, even though \geff defined in this work has only been validated in the regime $1\leq\geff\leq 150$, the phase transition is well described by this effective coupling value. 
This indicates that \geff captures the actual physical coupling of Yukawa systems up to the phase transition. 

At the transition itself, one expects a sudden change in \gmax as was observed in the 
two-dimensional case~\cite{Ott2011b} which signifies the 
re-ordering of the system into a body-centered cubic (bcc) crystal. A similar behavior is expected to appear at the transition from the bcc phase to the face-centered cubic which occurs in Yukawa systems at higher screening~\cite{Hamaguchi1997,Chugunov2003}. These questions are beyond the present analysis and will be studied elsewhere.

\begin{figure*}[t]
\includegraphics[width=7.85cm]{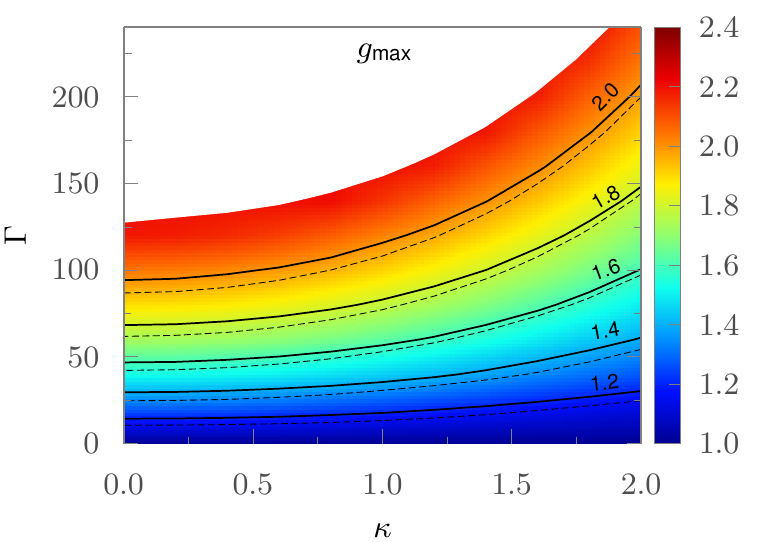}\hspace{1cm}
\includegraphics[width=7.85cm]{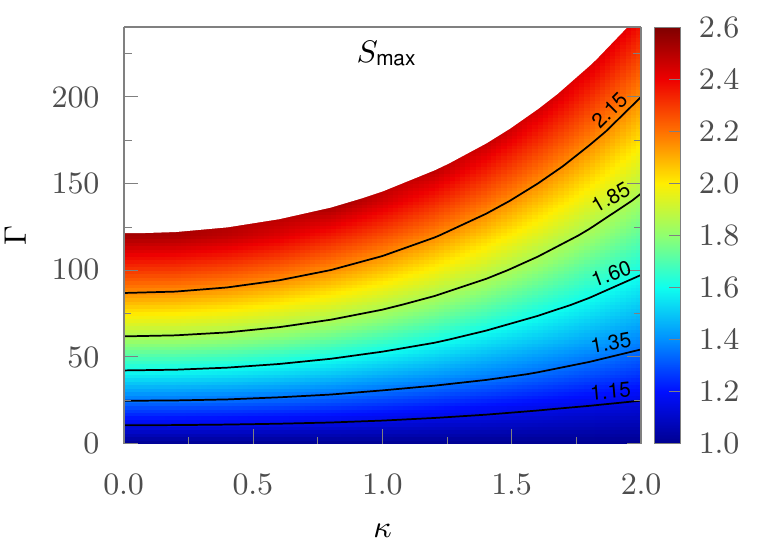}
\caption{HNC calculations for \gmax (left) and \smax (right) in the $\Gamma$-$\kappa$-plane. The contour lines of \smax have been overlayed onto \gmax in the left graph as broken lines. \label{fig:hnc}}
\end{figure*}

In addition, it is well known that the phase transition is closely connected to the properties of the static structure factor $S(k)$. More precisely, the Hansen-Verlet criterion~\cite{Hansen1969} states that the phase transition occurs
when the maximum peak \smax of $S(k)$ exceeds a threshold value of typically $2.85$. In order to connect the Hansen-Verlet criterion with the short-range definition of \geff at hand, we have performed additional
calculations in the Hypernetted Chain Approximation (HNC) to obtain both \smax and \gmax as a function of $\Gamma$ and $\kappa$. Figure~\ref{fig:hnc} shows the results of these calculations. While the HNC calculations do not extend 
all the way to the phase transition, one can see that both \gmax and \smax depend in  a qualitatively identical way on $\Gamma$. This is not a trivial result since $S(k)$ is related to an integral 
over $g(r)$ and thus depends on the complete $r$-dependence of the pair distribution function. We conclude that the phase transition is indicated by a critical value of $\gmax$ (and thus of \geff) in the same way 
as it is indicated by a critical value of \smax by the Hansen-Verlet criterion.

\section{Summary}
Particle correlations are a central issue in a wide range of plasma conditions and experiments. Despite the field's growing importance, 
there is lack of a clear, unified language when talking about the degree of correlation or the strength of coupling. In this work, we have proposed an approach 
based on the static structure of the system to define the degree of correlation as well as a simple way to measure the system's correlation from its structural properties. 
Our methodology is applicable to both unscreened, pure Coulomb systems as well as screened Yukawa systems with a Debye length corresponding to $\kappa\leq 2$, which encompasses 
almost all situations of interest. 

From an experimentalist's point of view, with the approach presented here, it suffices to have knowledge of the particle density $n$ and the radial pair distribution function (or the static structure function) to 
infer the coupling parameter $\Gamma$, instead of the measurement of the particle charge $Q$, the kinetic temperature $T$ and the particle density $n$. 

An assessment of the 
coupling strength based on the structure of the system furthermore allows one to make meaningful comparisons between Coulomb and Yukawa systems and between Yukawa systems with 
different screening. The common denominator is the effective coupling parameter \geff, which corresponds to the equivalent value of $\Gamma$ for a Coulomb system with the 
most similar nearest-neighbor structure. We have derived a definition of this effective coupling parameter \geff by considering the structural features of the respective systems 
during all stages of correlation build-up. Our definition~\eqref{eq:geff} is thus valid for the whole range of the strongly coupled liquid.

We also briefly remark on the need of knowing the screening length in order to apply our method to Yukawa systems. It is, in principle, possible to infer the value 
of the screening length non-invasively from the dynamics of the system, in much the same way as we have inferred the value of the coupling strength from the statics of the system (see Ref.~\cite{Ott2011b} 
for two-dimensional systems). Other methods include direct plasma measurements or the observation of self-excited waves~\cite{Nunomura2000,Nunomura2002}. The assumptions in this work, therefore, 
pose no principal limitation on the applicability of the methodology presented. 

The approach presented here for a one-component plasma is directly extendable to multicomponent plasmas. A detailed analysis of this generalization will be subject of a forthcoming paper.

\begin{acknowledgements}
This work is supported by the Deutsche Forschungsgemeinschaft via SFB-TR 24 (project A7) and the North-German Supercomputing Alliance (HLRN) via grant shp00006.
\end{acknowledgements}


\end{document}